\documentclass[twocolumn,showpacs,amsmath,nofootinbib,amssymb]{revtex4}

\usepackage{graphicx}% Include figure files
\usepackage{dcolumn}% Align table columns on decimal point
\usepackage{bm}% bold math

\newcommand{\Eqn}[1]{&\hspace{-0.2em}#1\hspace{-0.2em}&}

\begin{document}

\title{Energy conditions in modified Gauss-Bonnet gravity}

\author{Nadiezhda Montelongo Garc\'ia}
\email{nadiezhda@cosmo.fis.fc.ul.pt}\affiliation{Centro de Astronomia
e Astrof\'{\i}sica da Universidade de Lisboa, \\Campo Grande, Ed. C8
1749-016 Lisboa, Portugal}

\author{Tiberiu Harko}
\email{harko@hkucc.hku.hk} \affiliation{Department of Physics and
Center for Theoretical and Computational Physics, The University
of Hong Kong, Pok Fu Lam Road, Hong Kong}
\author{Francisco S. N. Lobo}
\email{flobo@cii.fc.ul.pt} \affiliation{Centro de Astronomia e
Astrof\'{\i}sica da Universidade de Lisboa, Campo Grande, Ed. C8
1749-016 Lisboa, Portugal}
\author{Jos\'e P. Mimoso}
\email{jpmimoso@cii.fc.ul.pt}\affiliation{Centro de Astronomia e
Astrof\'{\i}sica da Universidade de Lisboa, Campo Grande, Ed. C8
1749-016 Lisboa, Portugal}

\date{\today}

\begin{abstract}

In considering alternative higher-order gravity theories, one is liable to
be motivated in pursuing models consistent and inspired by several
candidates of a fundamental theory of quantum gravity. Indeed, motivations from string/M-
theory predict that scalar field couplings with the Gauss- Bonnet invariant, G, are important 
in the appearance of non-singular early time cosmologies. In this work, we discuss the 
viability of an interesting alternative gravitational theory, namely, modified Gauss-Bonnet 
gravity or $f(G)$ gravity. We consider specific realistic forms of $f(G)$ analyzed in the 
literature that account for the late-time cosmic acceleration and that have been found to cure 
the finite-time future singularities present in the dark energy models. We present the general 
inequalities imposed by the energy conditions and use the recent estimated values of the 
Hubble, deceleration, jerk and snap parameters to examine the viability of the above-mentioned 
forms of $f(G)$ imposed by the weak energy condition.

\end{abstract}


\maketitle

%%%%%%%%%%%%%%%%%%%%%%%%
\section{Introduction}
%%%%%%%%%%%%%%%%%%%%%%%%

An interesting approach in explaining the late-time accelerated expansion of the Universe \cite{expansion} is the possibility that at large scales Einstein's theory of General Relativity may break down. In this context, the Einstein field equation was first derived from an action principle by Hilbert, by adopting a linear function of the scalar curvature in the gravitational Lagrangian density. However, one may generalize the latter approach by considering higher order curvature invariants in the gravitational Lagrangian density \cite{fRgravity,early}. The motivation for this procedure consists in the analysis of strong gravitational fields near curvature singularities and in considering consistent candidates of a fundamental theory of quantum gravity. Indeed, string/M-theory predicts that scalar field couplings with the Gauss-Bonnet invariant $G$ are important in the appearance of non-singular early time cosmologies. These motivations may also be considered in the context of the late-time cosmic acceleration \cite{Nojiri:2005vv,modGB1,modGB2,Sotiriou:2007yd}).

An interesting alternative theory is modified Gauss-Bonnet gravity, or $f(G)$ gravity, where $f(G)$ is a general function of the Gauss-Bonnet term \cite{Zerbini, Nojiri:2005jg, F(G)-gravity,Nojiri:2007bt}. Note that the linear Gauss-Bonnet term is a topological invariant and the variation of the density $\sqrt{-g}G$ in the action leads to a total divergence, and therefore does not contribute to the field equations. Specific realistic models of $f(G)$ gravity were constructed to account for the late-time cosmic acceleration \cite{Nojiri:2007bt,Odintsov-models}, and it is these forms of $f(G)$ that we consider in this work. The respective constraints of the parameters of the models were also analyzed in \cite{Odintsov-models}.
More specifically, in \cite{Odintsov-models} the four types of finite-time future singularities emerging in the late-time accelerating era were studied from $f(G)$ gravity. It was shown that by taking into account higher-order curvature corrections the finite-time future singularities in $f(G)$ gravity are cured. Therefore, it turns out that adding such a non-singular modified gravity to singular dark energy models makes the combined theory to be non-singular one as well. In this context, we further consider the constraints imposed by the energy conditions and verify whether the parameter range of the specific models considered in \cite{Odintsov-models} are consistent with the energy conditions. More specifically, we define generalized energy conditions for $f(G)$ modified theories of gravity, and consider their realization for flat Friedmann cosmological models. In particular, we analyze whether the weak energy condition is satisfied by particular choices of $f(G)$ which were advocated in Refs. \cite{Nojiri:2007bt,Odintsov-models} as leading to viable models.

The energy conditions are fundamental to the singularity theorems and theorems of classical black hole thermodynamics (we refer the reader to \cite{hawkingellis} for more details).
Note that the energy conditions are obtained when one refers back to the Raychaudhuri equation for expansion, where the attractive character of gravity is reflected through the positivity condition, i.e., $R_{\mu\nu}k^\mu k^\nu \geq 0$, with $R_{\mu\nu}$ the Ricci tensor and $k^\mu$ any null vector. Now, in general relativity, through the Einstein field equation one ends up with $T_{\mu\nu}k^\mu k^\nu \geq 0$, which is the null energy condition.
In particular, the weak energy condition (WEC) assumes that the local energy density is positive and states that $T_{\mu\nu}U^\mu U^\nu \geq 0$, for all timelike vectors $U^\mu$, where $T_{\mu\nu}$ is the stress energy tensor (for a perfect fluid we have $\rho>0$ and $\rho+p\geq0$). By continuity, the WEC implies the null energy condition (NEC), $T_{\mu\nu}k^\mu k^\nu \geq 0$, where $k^\mu$ is a null vector~\cite{hawkingellis}.
The energy conditions have been extensively analyzed in the literature, such as in the cosmology settings and $f(R)$ gravity and we refer the reader to Refs. \cite{energyconditions,Santos:2007bs} for more details.

This paper is outlined in the following manner: In Section \ref{ref:II}, we present the gravitational field equations for modified Gauss Bonnet gravity, and in Section \ref{ref:III},  we outline the respective inequalities from the energy conditions. In Section \ref{ref:IV}, we consider specific forms of $f(G)$, and analyze the constraints arising from the energy conditions. Finally, in Section \ref{ref:conclusion} we present our conclusions. Throughout this work, we consider the following units $c={\cal G}=1$ (here ${\cal G}$ is the Newtonian gravitational constant to distinguish it from the Gauss-Bonnet term, $G$).

\section{Field equations of $f(G)$ modified gravity}\label{ref:II}

%%%%%%%%%%%%%%%%%%%%%%%%%%%%%%%%%%
%\subsection{Gravitational field equations}
%%%%%%%%%%%%%%%%%%%%%%%%%%%%%%%%%%

An interesting alternative gravitational theory is modified
Gauss-Bonnet gravity, which is given by the following action:
\begin{equation}
S=\frac{1}{2\kappa^2}\int d^4x \sqrt{-g}\left[R+f(G)\right]+S_M(g^{\mu\nu},\psi)\,,
   \label{modGBaction}
\end{equation}
where the Gauss-Bonnet invariant is defined as
\begin{equation}
G\equiv
R^2-4R_{\mu\nu}R^{\mu\nu}+R_{\mu\nu\alpha\beta}
R^{\mu\nu\alpha\beta}\,.
   \label{GBinvariant}
\end{equation}
It is also important to note that in the matter action, matter is
minimally coupled to the metric and not to the scalar field,
making Gauss-Bonnet gravity a metric theory. Thus, using the
diffeomorphism invariance of $S_M(g^{\mu\nu},\psi)$ yields the
covariant conservation of the stress-energy tensor, $\nabla^\mu
T_{\mu\nu}^{(mat)}=0$. Modified $f(G)$ gravity has been extensively analyzed in the literature and instead of reviewing all of its intricate details here, we refer the reader to \cite{modGB1,modGB2,Sotiriou:2007yd,Nojiri:2005jg},

Now varying the action (\ref{modGBaction}) with respect to the metric provides the following gravitational field equation
\begin{widetext}
\begin{eqnarray}
R_{\mu\nu}-\frac{1}{2}R g_{\mu\nu}
&=& \kappa^2 T^{(\mathrm{mat})}_{\mu \nu}
+\frac{1}{2}g_{\mu\nu} f(G)
%\nonumber \\
%&& \hspace{-30mm}
+\bigl(-2RR_{\mu\nu} +4R_{\mu\rho}R_{\nu}{}^{\rho}
-2R_{\mu}{}^{\rho\sigma\tau}R_{\nu\rho\sigma\tau}
+4g^{\alpha\rho}g^{\beta\sigma}R_{\mu\alpha\nu\beta}R_{\rho\sigma}
\bigr)f'(G)
\nonumber \\
&&
+2\left[{\nabla}_{\mu}{\nabla}_{\nu} f'(G) \right]R
-2g_{\mu \nu}\left[\Box f'(G) \right]R
+4\left[\Box f'(G) \right]R_{\mu \nu}
-4\left[{\nabla}_{\rho}{\nabla}_{\mu} f'(G) \right]
R_{\nu}{}^{\rho}
\nonumber \\
&&
-4\left[{\nabla}_{\rho}{\nabla}_{\nu} f'(G) \right]
R_{\mu}{}^{\rho}
+4g_{\mu \nu}\left[{\nabla}_{\rho}{\nabla}_{\sigma}
f'(G) \right]R^{\rho\sigma}
-4\left[{\nabla}_{\rho}{\nabla}_{\sigma} f'(G) \right]
g^{\alpha\rho}g^{\beta\sigma}R_{\mu\alpha\nu\beta}\,,
\label{eq:2.3}
\end{eqnarray}
\end{widetext}
where the prime denotes differentiation with respect to $G$.
Note that ${\nabla}_{\mu}$ is the covariant derivative operator associated with $g_{\mu \nu}$, $\Box \equiv g^{\mu \nu} {\nabla}_{\mu} {\nabla}_{\nu}$ is the covariant d'Alembertian, and $T^{(\mathrm{mat})}_{\mu \nu}$ is the contribution to the stress energy tensor from ordinary matter.

In this paper, we consider the flat FRW space-time described by the metric
\begin{equation}
ds^{2}=-dt^{2}+a^{2}(t)d \mathbf{x}^{2}\,,
\label{metric}
\end{equation}
where $a(t)$ is the scale factor.

In the FRW background, and taking into account a perfect fluid equation of state for ordinary matter, it follows that the field equations for $f(G)$ gravity are given by
\begin{eqnarray}
&& \hspace{-5mm} 24H^{3}\dot{f}'(G)+6H^{2}+f(G)-G
f'(G)=2\kappa^{2}\rho\,,
\label{dancke} \\
&& \hspace{-5mm}
8H^{2}\ddot{f}'(G)+16H\dot{f}'(G)\left(\dot{H}+H^{2}\right)
+\left(4\dot{H}+6H^{2}\right)
    \nonumber  \\
&&+f(G)-G f'(G)=-2\kappa^{2}p\,, \label{bitte}
\end{eqnarray}
where $\rho$ and $p$ are the energy density and pressure, respectively, and the overdot denotes a derivative with respect to the time coordinate, $t$.

Moreover, we have
\begin{eqnarray}
R \Eqn{=} 6 \left(2H^{2}+\dot H \right)\,,
\label{eq:R} \\
G \Eqn{=} 24H^{2} \left( H^{2}+\dot H \right)\,. \label{eq:G}
\end{eqnarray}

In the FRW background, the gravitational field equations may be rewritten to take the following form
\begin{equation}
\rho_{\mathrm{eff}}=\frac{3}{\kappa^{2}}H^{2}\,,
\quad
p_{\mathrm{eff}}=-\frac{1}{\kappa^{2}} \left( 2\dot H+3H^{2} \right)\,,
\label{GutenTag}
\end{equation}
where $\rho_{\mathrm{eff}}$ and $p_{\mathrm{eff}}$ are
the effective energy density and pressure, respectively, defined as
\begin{widetext}
\begin{eqnarray}
\rho_{\mathrm{eff}} \Eqn{=} \frac{1}{2\kappa^{2}} \left[
-f(G)+24H^{2} \left(H^{2}+\dot H\right)f'(G)-24^{2}H^{4}
\left(2\dot H^{2}+H\ddot H+4H^{2}\dot H\right)f''(G) \right]
 +\rho\,, \label{eq:rho-eff-1}
\\
p_{\mathrm{eff}} \Eqn{=} \frac{1}{2\kappa^{2}}\Bigl\{f(G)-24H^{2}
\left(H^{2}+\dot H\right)f'(G) +(24)8H^{2}\Big[ 6\dot{H}^{3}+8 H
\dot{H} \ddot{H} + 24\dot{H}^{2} H^2 + 6H^3\ddot{H}
\nonumber\\
& & + 8H^4\dot{H}+H^{2} \dddot{H} \Big]f''(G)+8(24)^{2}H^{4}
\left(2\dot{H}^2+H\ddot{H}+4H^{2}\dot{H}\right)^{2}f'''(G)\Bigr\}+p\,,
\label{eq:p-eff-1}
\end{eqnarray}
where Eqs.~(\ref{eq:R})-(\ref{eq:G}) were used.

We also present the following useful relationship
\begin{eqnarray}
\nonumber
\rho_{\mathrm{eff}}+p_{\mathrm{eff}} &=& \rho+p + \frac{96H^2}{\kappa^{2}}\Bigl[\Big( 6\dot{H}^{3}+8 H\dot{H} \ddot{H} - 18\dot{H}^{2} H^2 + 3H^3\ddot{H}- 4H^4\dot{H}+H^{2} \dddot{H} \Big)f''(G)
\nonumber\\
& & +24H^{2}
\left(2\dot{H}^2+H\ddot{H}+4H^{2}\dot{H}\right)^{2}f'''(G)\Bigr]\,,
\label{eq:NEC}
\end{eqnarray}
as it will be used throughout the text in the context of the energy conditions.

\end{widetext}

%%%%%%%%%%%%%%%%%%%%%%%%%%%%%
\section{Energy Conditions}\label{ref:III}
%%%%%%%%%%%%%%%%%%%%%%%%%%%%%

%%%%%%%%%%%%%%%%%%%%%%%%%%%%%%%%%%
%\subsection{General Relativity}
%%%%%%%%%%%%%%%%%%%%%%%%%%%%%%%%%%

The energy conditions arise when one refers to the Raychaudhuri equation for the expansion, given by
\begin{equation}
\label{Raych}
\frac{d\theta}{d\tau}= - \frac{1}{2}\,\theta^2 -
\sigma_{\mu\nu}\sigma^{\mu\nu} + \omega_{\mu\nu}\omega^{\mu\nu}
- R_{\mu\nu}k^{\mu}k^{\nu} \;,
\end{equation}
where $R_{\mu\nu}$  is the Ricci tensor, and $\theta\,$, $\sigma^{\mu\nu}$ and $\omega_{\mu\nu}$ are, respectively, the expansion, shear and rotation associated to the congruence defined by the null vector field $k^{\mu}$. Note that the Raychaudhuri equation is a purely geometric statement, and as such it makes no reference to any gravitational field equations.

The shear is a ``spatial" tensor with $\sigma^2 \equiv \sigma_{\mu\nu}\sigma^{\mu\nu}\geq 0$, thus from Raychaudhury's equation it is clear that for any hypersurface orthogonal congruences, which imposes $\omega_{\mu\nu} \equiv 0$, the condition for attractive gravity reduces to $R_{\mu\nu}k^{\mu}k^{\nu}\geq 0$. The latter inequality ensures that geodesic congruences focus within a finite value of the parameter labeling points on the geodesics. However, in general relativity, through the
Einstein field equation one can write the above condition in terms
of the stress-energy tensor given by $T_{\mu\nu}k^\mu k^\nu \ge
0$. In any other theory of gravity, one would require to know how
one can replace $R_{\mu\nu}$ using the corresponding field
equations. In particular, in a theory where we still have an Einstein-Hilbert term, the task of evaluating $R_{\mu\nu}k^\mu k^\nu$ is trivial. However, in $f(G)$ modified theories of gravity under consideration, things are not so straightforward.

For convenience Eq. (\ref{eq:2.3}) may be written as the following effective gravitational field equation
\begin{equation}
G_{\mu\nu}\equiv R_{\mu\nu}-\frac{1}{2}R\,g_{\mu\nu}= T^{{\rm
eff}}_{\mu\nu} \,,
    \label{field:eq2}
\end{equation}
where the effective stress-energy tensor is given by
\begin{widetext}
\begin{eqnarray}
T^{{\rm eff}}_{\mu\nu}
&=& \kappa^2 T^{(\mathrm{mat})}_{\mu \nu}
+\frac{1}{2}g_{\mu\nu} f(G)
%\nonumber \\
%&& \hspace{-30mm}
+\bigl(-2RR_{\mu\nu} +4R_{\mu\rho}R_{\nu}{}^{\rho}
-2R_{\mu}{}^{\rho\sigma\tau}R_{\nu\rho\sigma\tau}
+4g^{\alpha\rho}g^{\beta\sigma}R_{\mu\alpha\nu\beta}R_{\rho\sigma}
\bigr)f'(G)
\nonumber \\
&&
+2\left[{\nabla}_{\mu}{\nabla}_{\nu} f'(G) \right]R
-2g_{\mu \nu}\left[\Box f'(G) \right]R
+4\left[\Box f'(G) \right]R_{\mu \nu}
-4\left[{\nabla}_{\rho}{\nabla}_{\mu} f'(G) \right]
R_{\nu}{}^{\rho}
\nonumber \\
&&
-4\left[{\nabla}_{\rho}{\nabla}_{\nu} f'(G) \right]
R_{\mu}{}^{\rho}
+4g_{\mu \nu}\left[{\nabla}_{\rho}{\nabla}_{\sigma}
f'(G) \right]R^{\rho\sigma}
-4\left[{\nabla}_{\rho}{\nabla}_{\sigma} f'(G) \right]
g^{\alpha\rho}g^{\beta\sigma}R_{\mu\alpha\nu\beta}\,.
\label{efffield2}
\end{eqnarray}
\end{widetext}

In this context, the positivity condition, $R_{\mu\nu}k^\mu k^\nu \geq 0$, in the Raychaudhuri equation provides the following form for the null
energy condition $T^{{\rm eff}}_{\mu\nu} k^\mu k^\nu\geq 0$,
through the modified gravitational field equation
(\ref{field:eq2}). We also impose the condition $T^{({\rm mat})}_{\mu\nu} k^\mu k^\nu\ge 0$ for ordinary matter. This is useful as applying local Lorentz transformations it is possible to show that the above condition implies that the energy density is positive in all local frames of reference.

Taking into account that the Raychaudhuri equation holds for any geometrical theory of gravitation, we will maintain its physical motivation, namely, the focussing of geodesic congruences, along with the attractive character of the gravitational interaction to deduce the energy conditions in the context of
$f(G)$ modified gravity. To this end, using the modified (effective) gravitational field equations the energy conditions in this context are given by
\begin{equation}
{\rm NEC} \Longleftrightarrow \rho_{\mathrm{eff}} +p_{\mathrm{eff}} \geq 0 \,,
  \label{NEC}
\end{equation}
\begin{equation}
{\rm WEC} \Longleftrightarrow \rho_{\mathrm{eff}} \geq 0 \; {\rm and} \; \rho_{\mathrm{eff}} +p_{\mathrm{eff}} \geq 0 \,,
   \label{WEC}
\end{equation}
\begin{equation}
{\rm SEC} \Longleftrightarrow \rho_{\mathrm{eff}} +3p_{\mathrm{eff}} \geq 0 \; {\rm and} \; \rho_{\mathrm{eff}} +p_{\mathrm{eff}} \geq 0 \,,
   \label{SEC}
\end{equation}
\begin{equation}
{\rm DEC} \Longleftrightarrow \rho_{\mathrm{eff}} \geq 0 \; {\rm and} \; \rho_{\mathrm{eff}} \pm p_{\mathrm{eff}} \geq 0 \,,
   \label{DEC}
\end{equation}
where the notation NEC, WEC, SEC and DEC stand for the null, weak, strong and dominant energy conditions, respectively.

%%%%%%%%%%%%%%%%%%%%%%%%%%%%%%%%%%
%\subsection{$f(G)$ modified gravity}
%%%%%%%%%%%%%%%%%%%%%%%%%%%%%%%%%%

Now, in standard mechanics terminology the first four time derivatives of position are referred to as velocity, acceleration, jerk and snap. In a cosmological setting, in addition to the Hubble parameter $H=\dot{a}/a$, it is appropriate to define the deceleration, jerk, and snap parameters as
\begin{equation}
q=-\frac{1}{H^2}\frac{\ddot{a}}{a}\;, \qquad j=\frac{1}{H^3}\frac{\dddot{a}}{a}\;,
\quad {\rm{and}} \quad s=\frac{1}{H^4}\frac{\ddddot{a}}{a}\;,
\end{equation}
respectively.

In terms of the these parameters, we consider the following definitions
\begin{eqnarray}
&\dot{H}=-H^2(1+q)\;, \\
&\ddot{H}=H^3(j+3q+2)\;, \\
&\dddot{H}=H^4(s-2j-5q-3)\;,
\end{eqnarray}
respectively.

Using the above definitions, then the energy conditions (\ref{NEC})-(\ref{DEC}) take the following respective forms
%\vspace{1cm}
\begin{widetext}
\begin{eqnarray}
\nonumber
{\rm NEC}:\qquad \rho_{\textrm{eff}}+p_{\textrm{eff}}=\rho+p+\frac{96}{k^{2}}\left\{ -(6q^3+27q^2+21q+8qj+9j-s)f''(G)+\right.\\
\left.24[4(q^{2}+2q+1)H^{2}+2q^{2}+7q+j+4]f'''(G)\right\}H^{8}\geq 0\,,
\end{eqnarray}
\begin{eqnarray}
{\rm WEC}:\qquad
\rho_{\textrm{eff}}=\rho+\frac{1}{2k^2}\left[-f(G)-24H^{4} q f'(G)-(24)^{2}H^8(2q^2+3q+j)f''(G) \right]\geq 0,~ \qquad\rho_{\textrm{eff}}+p_{\textrm{eff}}\geq 0\,,
\end{eqnarray}
\begin{eqnarray}
{\rm SEC}:\qquad \rho_{\textrm{eff}}+3p_{\textrm{eff}}&=&\rho+3p+\frac{1}{k^{2}}[f(G)+24H^4qf'(G)
+288H^{8}(-6q^3-23q^2-15q-8qj-7j+s)f''(G)\nonumber \\
&&+(24)(288)H^{12}(2q^2+3q+j)^2f'''(G)]\geq 0          ~,\qquad\rho_{\textrm{eff}}+p_{\textrm{eff}}\geq 0 \,,
\end{eqnarray}
\begin{eqnarray}
{\rm DEC}:\qquad\rho_{\textrm{eff}}-p_{\textrm{eff}}&=&\rho-p+\frac{1}{k^2}
[-f(G)-24H^{4}q f'(G)
-96H^{8}(-6q^3-15q^2-3q-8qj-3j+s)f''(G)\nonumber \\
&&-(24)(96)H^{12}(2q^2+3q+j)^{2}f'''(G)]\geq 0\,,
\qquad \rho_{\textrm{eff}}+p_{\textrm{eff}}\geq 0,\qquad \rho_{\textrm{eff}}\geq 0 \,.
\end{eqnarray}
\end{widetext}

%%%%%%%%%%%%%%%%%%%%%%%%%%%%%%%%%%%%%%%%
\section{Viable $f(G)$ theories using the energy conditions}\label{ref:IV}
%%%%%%%%%%%%%%%%%%%%%%%%%%%%%%%%%%%%%%%%

Viable $f(G)$ modified theories of gravity were used in \cite{Nojiri:2007bt} to account for the late-time cosmic acceleration. These latter models were studied in the context of curing the four types of finite-time future singularities emerging in the late-time accelerating era \cite{Odintsov-models}. Indeed it was shown that by taking into account higher-order curvature corrections, in the context of $f(G)$ gravity, the finite-time future singularities are cured. In this context, we further consider the constraints imposed by the energy conditions and verify whether the parameter range of the specific models considered in \cite{Odintsov-models} are consistent with the energy conditions for flat Friedman cosmological models.

Thus, we consider some specific forms of $f(G)$, considered in \cite{Nojiri:2007bt,Odintsov-models} given by
\begin{eqnarray}
f_{1}(G) \Eqn{=} \frac{a_{1}G^{n}+b_{1}}{a_{2}G^{n}+b_{2}}\,,
\label{uno} \\
%F_{2}(G) \Eqn{=} \frac{a_{1}G^{n+N}+b_{1}}{a_{2}G^{n}+b_{2}}\,,
%\label{due} \\
f_{2}(G) \Eqn{=} a_{3} G^{n}(1+b_{3} G^{m})\,,
\label{terzo}
\end{eqnarray}
where $a_{1}$, $a_{2}$, $b_{1}$, $b_{2}$, $a_{3}$, $b_{3}$, $n$, and $m$
are constants. In the following, we always assume $n > 0$.

Note that the Gauss-Bonnet invariant, defined in Eq. (\ref{eq:G}), can be expressed as
\begin{equation}
G=-24H^4 q \,,
\end{equation}
in terms of the Hubble and the deceleration parameters, respectively.

As the inequalities imposed by the energy conditions in $f(G)$ gravity are extremely lengthly, in the following analysis we only consider the WEC in exemplifying the application of the energy conditions. We consider the following present-day values for the deceleration, jerk and snap parameters \cite{Rap,Poplawski:2006ew}:
$q_0=-0.81\pm 0.14$, $j_0=2.16^{+0.81}_{-0.75}$, and $s_0=-0.22^{+0.21}_{-0.19}$.

%%%%%%%%%%%%%%%%%%%%%%%%%%%%%%%%%%%%%%%%
\subsection{Specific case: $ f_{1}(G) = \frac{a_{1}G^{n}+b_{1}}{a_{2}G^{n}+b_{2}}$}
%%%%%%%%%%%%%%%%%%%%%%%%%%%%%%%%%%%%%%%%

In first place, we consider the specific case of Eq. (\ref{uno}). For simplicity in the examples analyzed we consider vacuum, i.e., $\rho =p=0$. The WEC constraints, i.e., $\rho_{\textrm{eff}}\geq 0$ and $\rho_{\textrm{eff}}+p_{\textrm{eff}}\geq 0$, are respectively given by
\begin{widetext}
\begin{eqnarray}
\nonumber
&&-[a_{1}(-24qH^4)^n+b_{1}][a_{2}(-24qH^4)^n+b_{2}]+n(-24qH^{4})^n
(a_{1}b_{2}-a_{2}b_{1})+(24)^{2}nH^{8}(-24qH^{4})^{n-2}
[a_{2}(n+1)\times\\
&&\times(-24qH^{4})^{n}+b_{2}(1-n)]
\frac{(a_{1}b_{2}-a_{2}b_{1})(2q^2+3q+j)}{a_{2}(-24qH^4)^{n}+b_{2}}
\geq 0 \,, \label{NEC1a}
\end{eqnarray}
\begin{eqnarray}
\nonumber
&&n(-24qH^{4})^{n}(-b_{1}a_{2}
+a_{1}b_{2})\{(6q^{3}+27q^{2}+21q+8qj+9j-s)[n(a_{2}^{2}(-24qH^{4})^{2n}-b_{2}^{2})+
(a_{2}(-24qH^{4})^{n}+b_{2})^{2}]\\
\nonumber
&&+(4H^{2}+8H^{2}q+4H^{2}q^{2}+2q^{2}+7q+j+4)[4a_{2}b_{2}(1-n^{2})(-24qH^{4})^{n}+a_{2}^{2}(n^{2}+3n+2)(-24qH^{4})^{2n}
\\
&&+b_{2}^{2}(n^{2}-3n+2)]q^{-1}H^{-4}
\}\geq 0\,.\label{NEC1b}
\end{eqnarray}

\begin{figure}[ht]
  \centering
  \includegraphics[width=2.6in]{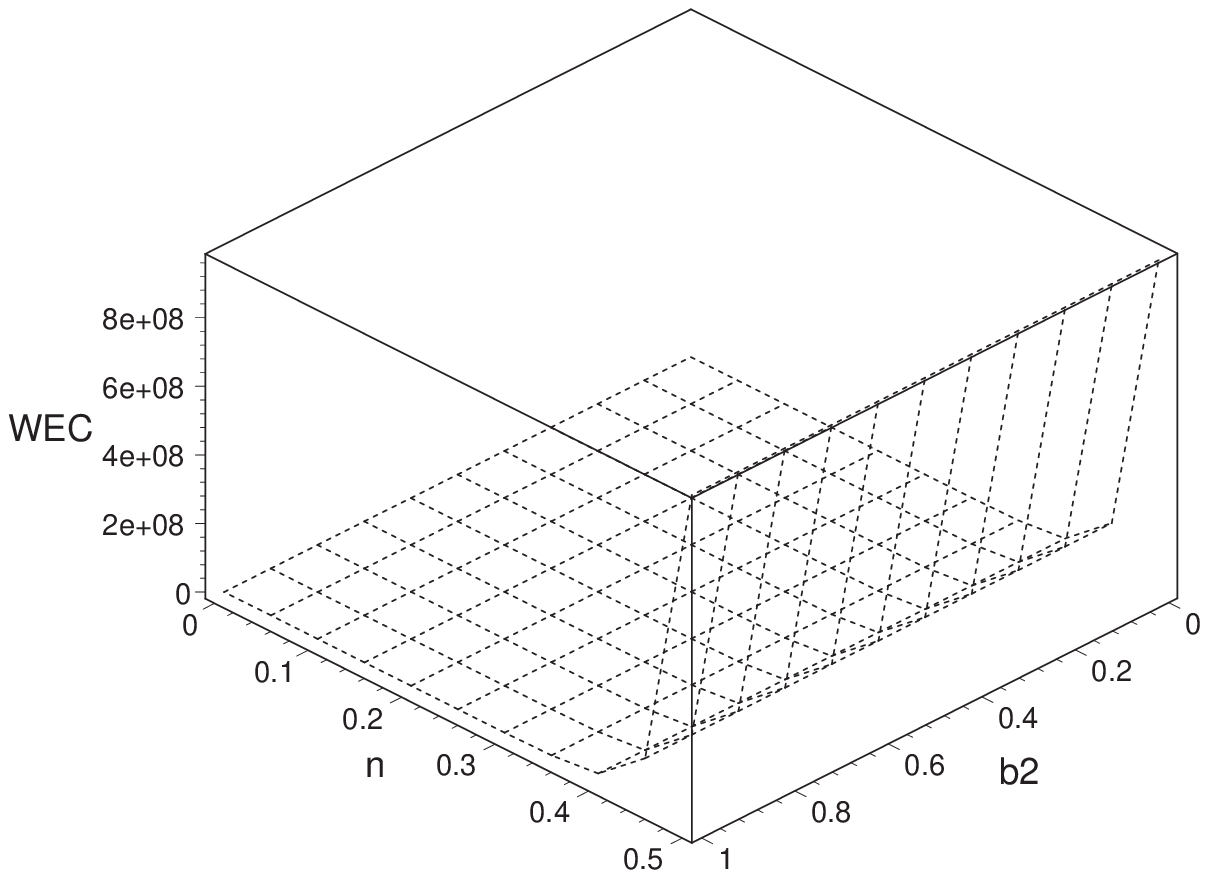}
   \includegraphics[width=2.6in]{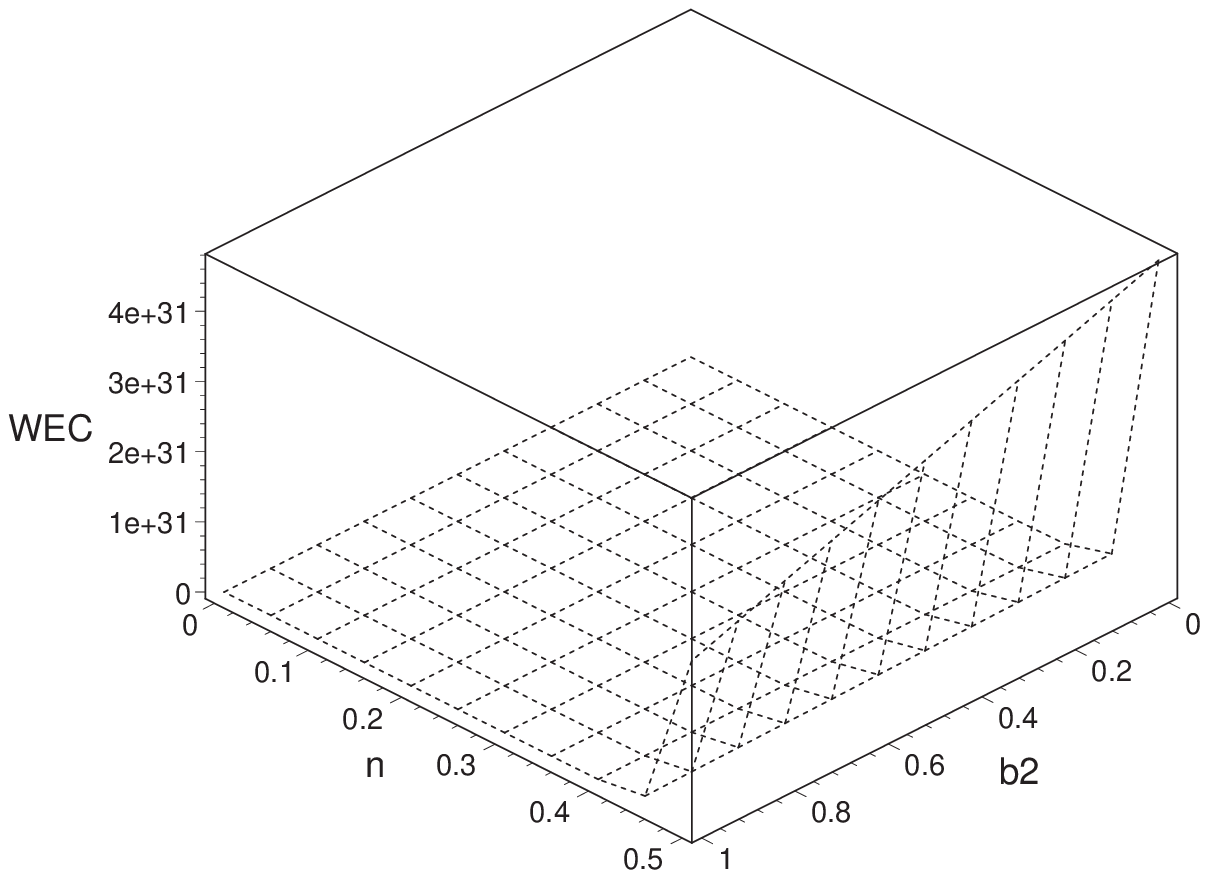}
  \caption{The plots depict the weak energy condition for the specific form of $ f_{1}(G) = \frac{a_{1}G^{n}+b_{1}}{a_{2}G^{n}+b_{2}}$. The left plot corresponds to $\rho_{\textrm{eff}}\geq 0$; the right plot corresponds to $\rho_{\textrm{eff}}+p_{\textrm{eff}}\geq 0$. We have considered the values $a_1=-1$, $b_1=-1$, and $a_2=2$. The plots show that the weak energy condition are satisfied for the parameter range considered. See the text for details.}
  \label{fig:WEC1a}
\end{figure}

\end{widetext}

The constraints provided by the inequalities (\ref{NEC1a})-(\ref{NEC1b}) are too complicated to find exact analytical expressions for the parameter ranges of the constants $a_{1}$, $a_{2}$, $b_{1}$, $b_{2}$, and $n$, so we consider specific values for some of the parameters. In particular, we impose the following values  $a_1=-1$, $b_1=-1$, and $a_2=2$, and plot the WEC as a function of $b_2$ and $n$, which is depicted in Fig. \ref{fig:WEC1a}. The latter does indeed prove that the specific form of $ f_{1}(G)$ given by Eq. (\ref{uno}) considered in \cite{Odintsov-models} is consistent with the WEC inequalities.

%%%%%%%%%%%%%%%%%%%%%%%%%%%%%%%%%%%%%%%%
\subsection{Specific case:
$f_{2}(G) = a_{3} G^{n}(1+b_{3} G^{m})$}
%%%%%%%%%%%%%%%%%%%%%%%%%%%%%%%%%%%%%%%%

\begin{figure*}[ht]
  \centering
  \includegraphics[width=2.6in]{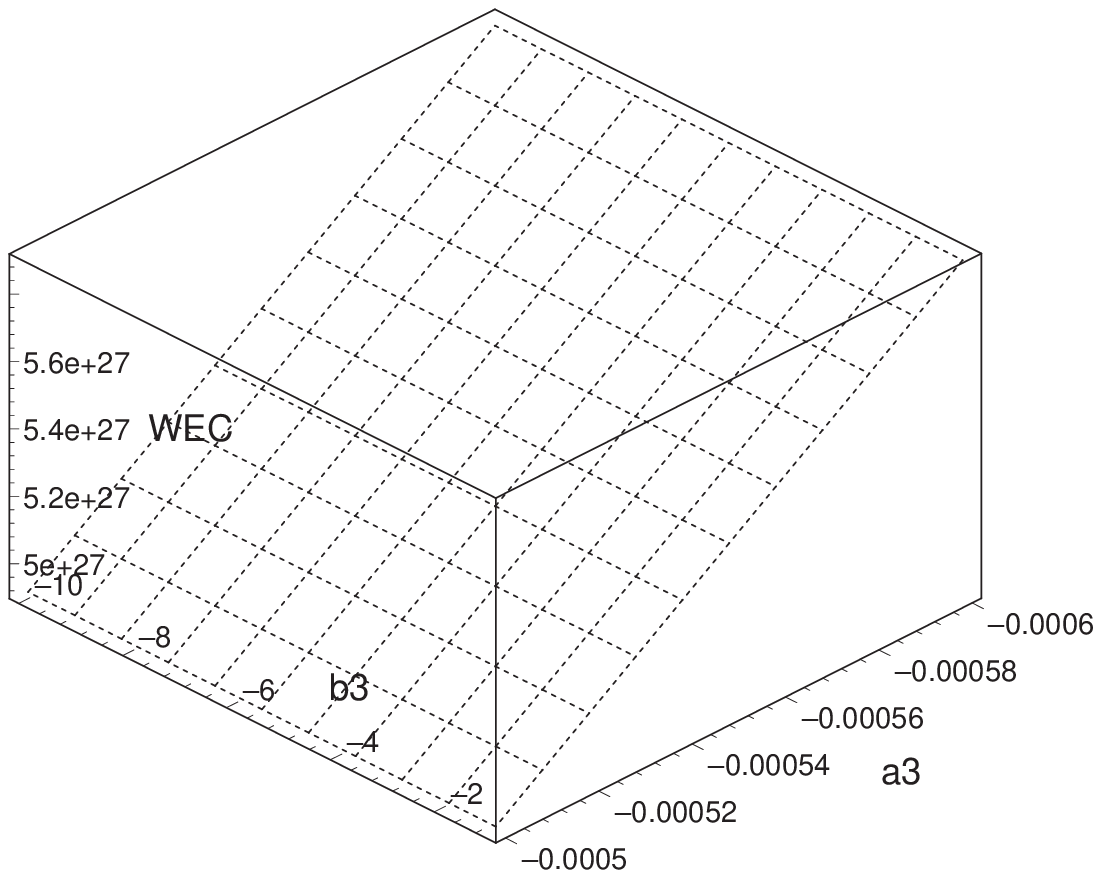}
   \includegraphics[width=2.6in]{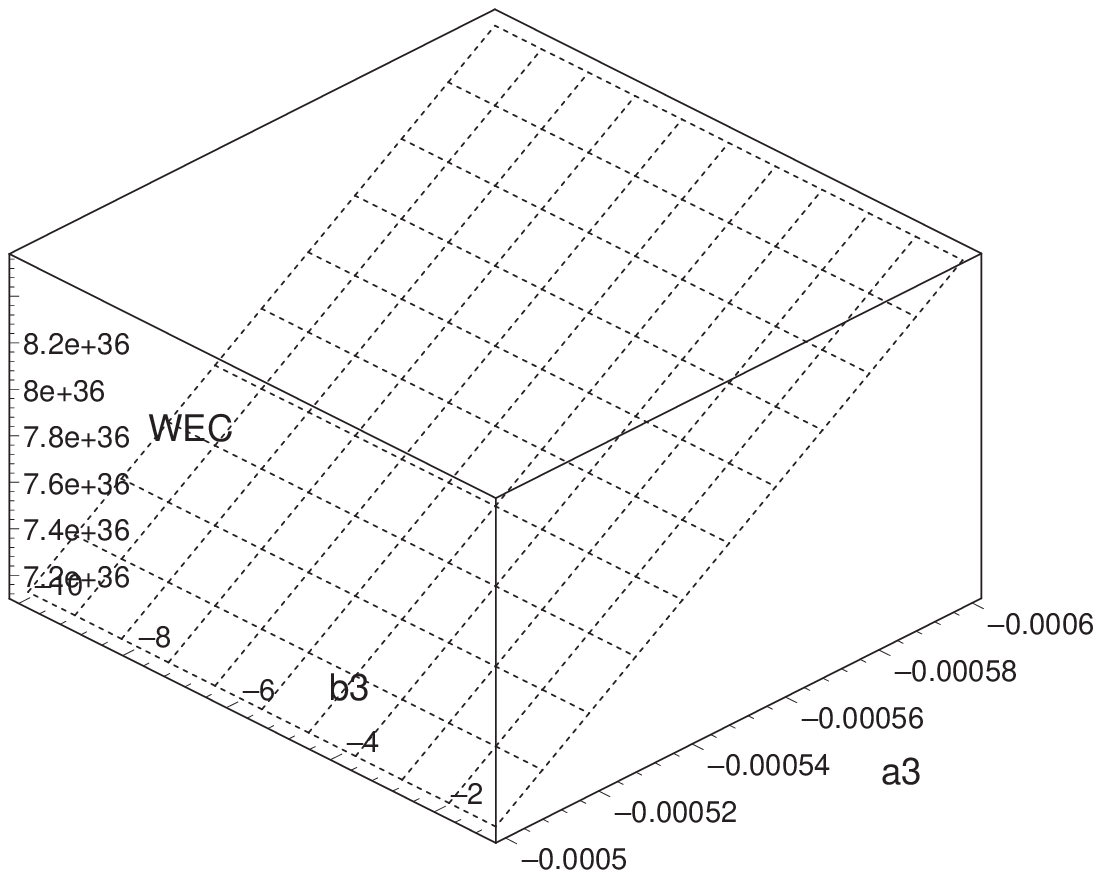}
  \caption{The plots depict the weak energy condition for the specific form of $f_{2}(G) = a_{3} G^{n}(1+b_{3} G^{m})$. The left plot corresponds to $\rho_{\textrm{eff}}\geq 0$; the right plot corresponds to $\rho_{\textrm{eff}}+p_{\textrm{eff}}\geq 0$. The parameter range for this specific case is given: $n>\frac{1}{2}, n\neq 1, m<0, n+m>1,a_{3}b_{3}>0$. We have considered the specific values $(n=2.5,m=-1)$. See the text for details.}
  \label{fig:WECB2}
\end{figure*}
\begin{figure*}[ht]
  \centering
  \includegraphics[width=2.6in]{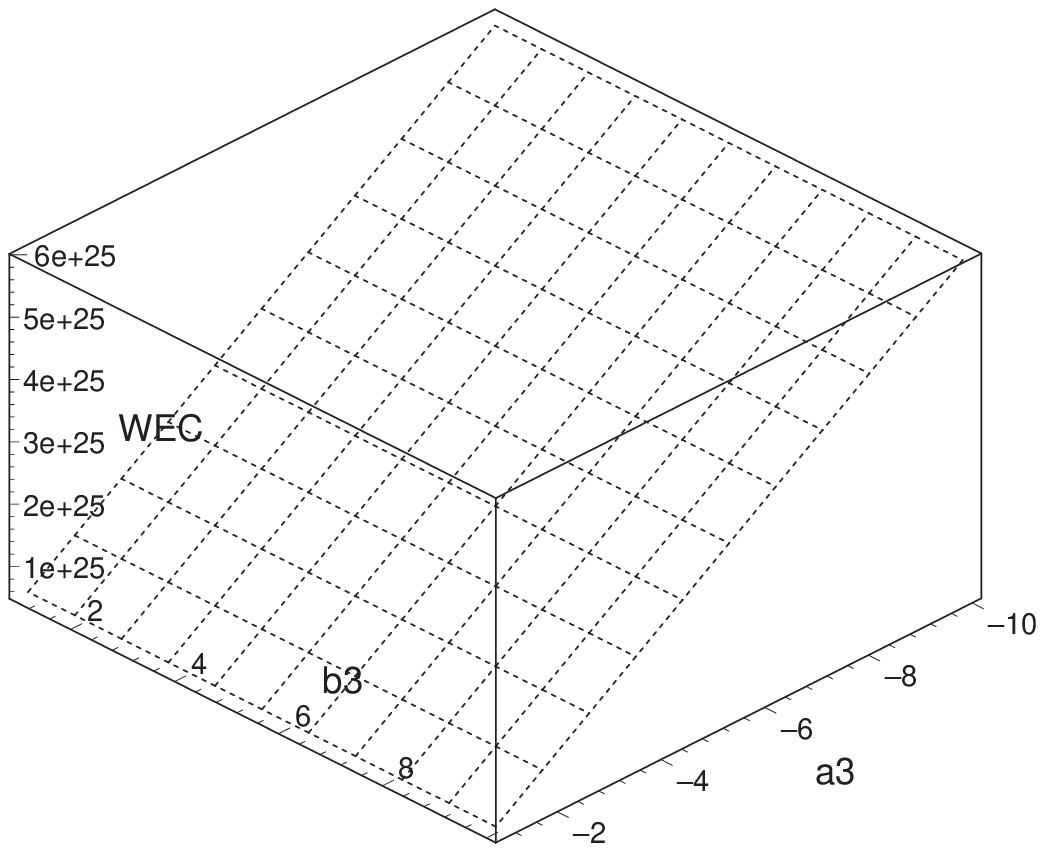}
   \includegraphics[width=2.6in]{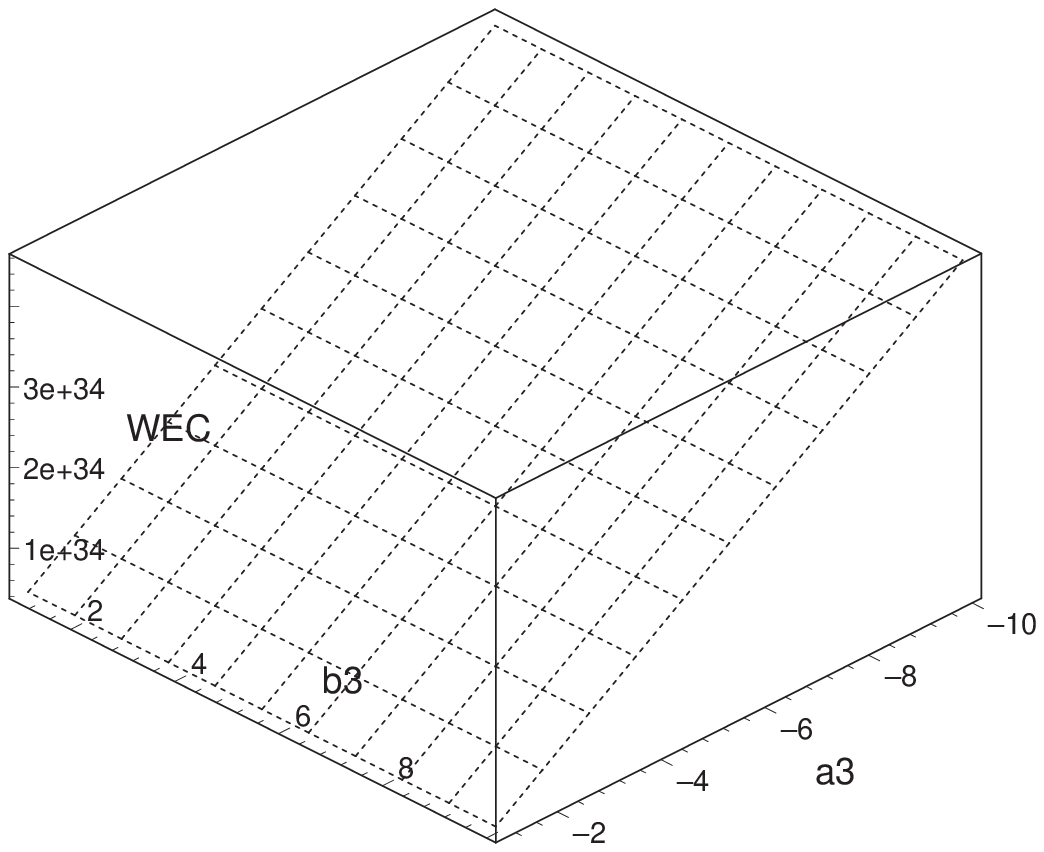}
  \caption{The plots depict the weak energy condition for the specific form of $f_{2}(G) = a_{3} G^{n}(1+b_{3} G^{m})$. The left plot corresponds to $\rho_{\textrm{eff}}\geq 0$; the right plot corresponds to $\rho_{\textrm{eff}}+p_{\textrm{eff}}\geq 0$. The parameter range for this specific case corresponds to: $n>\frac{1}{2}, n\neq 1,m<0,\frac{2}{3}<n+m<1,a_{3}b_{3}<0$. We have considered the specific values $(n=1.8,m=-1)$. See the text for details.}
  \label{fig:WECB3}
\end{figure*}
\begin{figure*}[ht]
  \centering
  \includegraphics[width=2.6in]{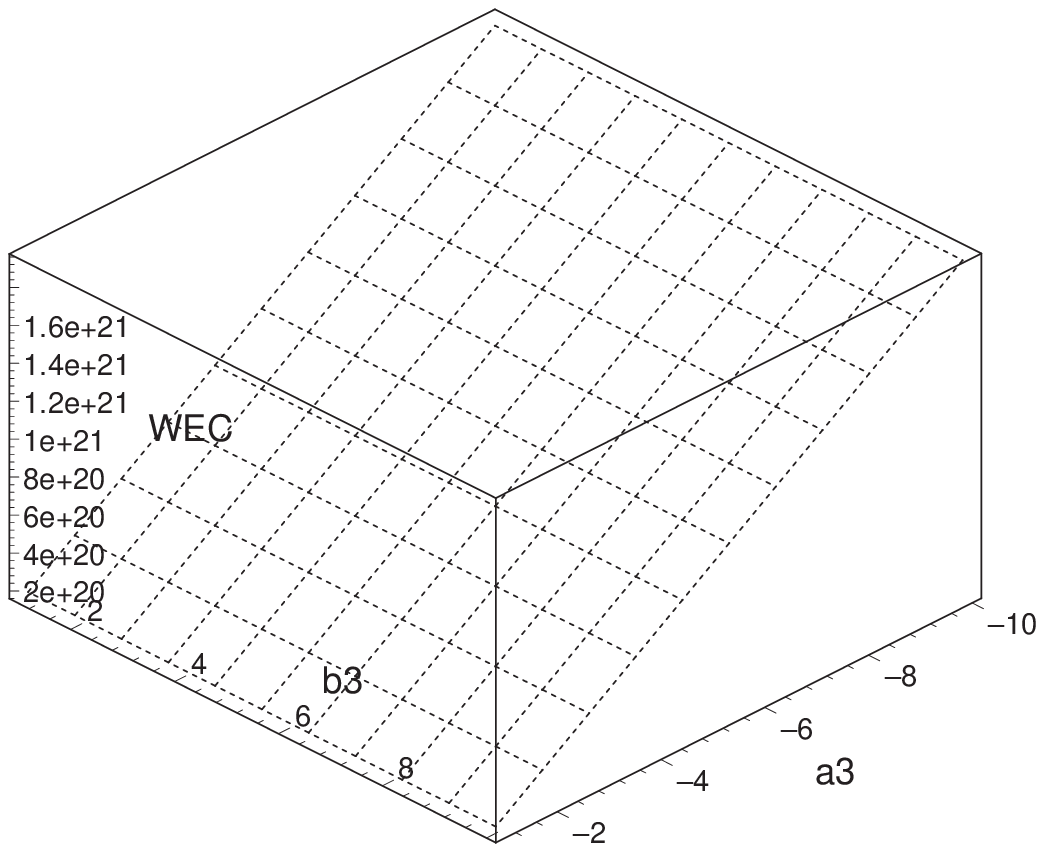}
   \includegraphics[width=2.6in]{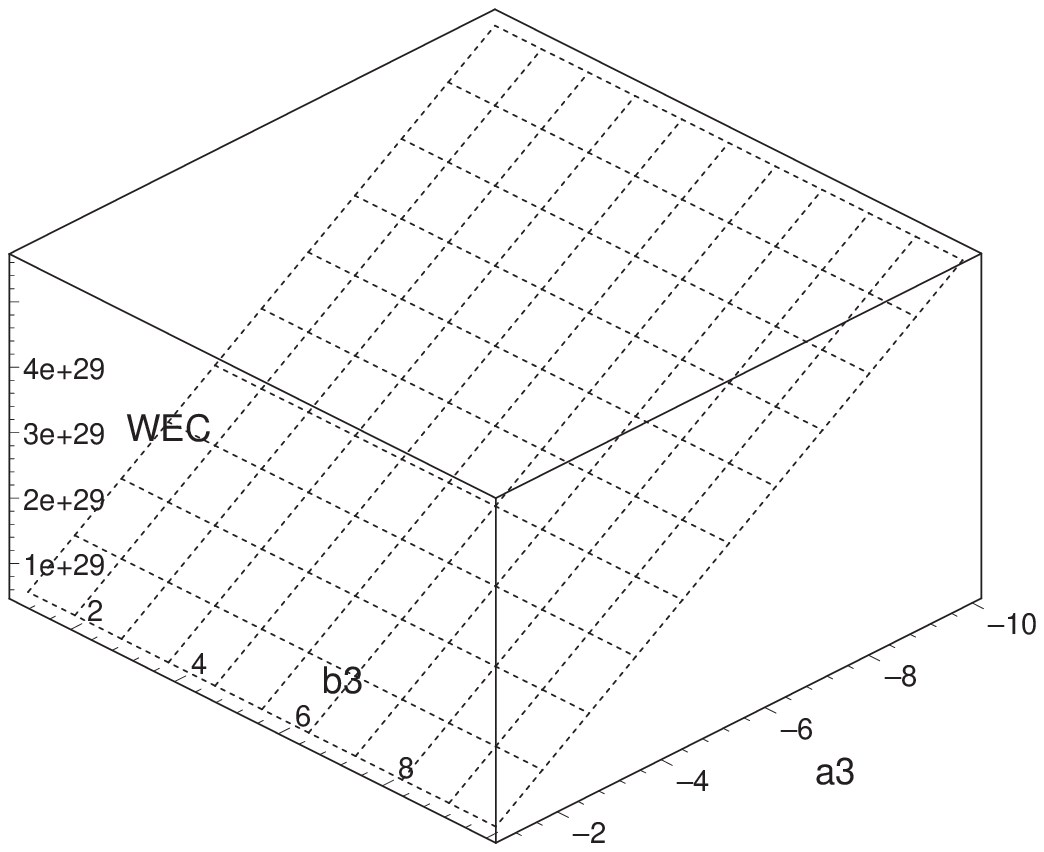}
  \caption{The plots depict the weak energy condition for the specific form of $f_{2}(G) = a_{3} G^{n}(1+b_{3} G^{m})$. The left plot corresponds to $\rho_{\textrm{eff}}\geq 0$; the right plot corresponds to $\rho_{\textrm{eff}}+p_{\textrm{eff}}\geq 0$. The parameter range for this specific case corresponds to: $n>\frac{1}{2}, n\neq 1, m<0, n+m<\frac{2}{3}$. We have considered the specific values of $n=1.3,m=-1$. See the text for details.}
  \label{fig:WECB4}
\end{figure*}

We consider the specific realistic case of Eq. (\ref{terzo}) analyzed in \cite{Odintsov-models} which accounts for the late-time cosmic acceleration, and that cured the four types of finite-time future singularities emerging in the late-time accelerating era, given by the following specific conditions $n>0$, $m<0$ and $n\neq 1$ and for several parameter ranges. Rather than exhaustively analyze all of the cases, we consider a specific case that does indeed prove that in addition to curing the finite-time future singularities is satisfies the weak energy condition. The latter parameter range is given by following
\begin{eqnarray}
&n>1/2,\quad n+m> 1 \quad {\rm and}\quad a_{3}b_{3}>0\,, \label{f2a}\\
&n>1/2, \quad 2/3<n+m < 1 \quad
{\rm and} \quad a_{3}b_{3}<0\,,\label{f2b}\\
&n>1/2, \quad {\rm and} \quad  n+m \leq 2/3  \label{f2c}\,.
\end{eqnarray}

For the form of $f_2(G)$ considered by Eq. (\ref{terzo}), the WEC constraints, i.e., $\rho_{\textrm{eff}}\geq 0$ and $\rho_{\textrm{eff}}+p_{\textrm{eff}}\geq 0$, are given by
\begin{widetext}
\begin{eqnarray}
\nonumber
&&-a_{3}\{(-24qH^4)^{n}[1+b_{3}(-24qH^{4})^{m}]+24H^{4}
(-24qH^{4})^{n-1}[n+(n+m)b_{3}(-24qH^{4})^{m}]\\
&&+24^{2}H^{8}(-24qH^{4})^{n-1}[n
+(n+m)b_{3}(-24qH^{4})^{m}](2q^{2}+3q+j)
\}\geq 0 \,,\label{NEC2a}
\end{eqnarray}
\begin{eqnarray}
&&a_{3}(-24qH^{4})^{n}\{(6q^{3}+27q^{2}+21q+8qj-s)[n^{2}-n+b_{3}(-24qH^{4})^{m}
(n^{2}-n+2nm+m^{2}-m)]
    \nonumber \\
\nonumber
&&+(4H^{2}+8H^{2}q+4H^{2}q^{2}+2q^{2}+7q+j+4)[n^{3}-3n^{2}+2n
+b_{3}(-24qH^{4})^{m}(n^{3}-3n^{2}+3n^{2}m\\
&&+2n-6nm+3nm^{2}+m^{3}-3m^{2}+2m)]q^{-1}H^{-4} \}\geq 0\,,\label{NEC2b}
\end{eqnarray}
\end{widetext}
respectively.

As in the previous example, the constraints provided by the inequalities (\ref{NEC2a})-(\ref{NEC2b}) are too complicated to find exact analytical expressions for the respective parameter ranges of the constants $a_{3}$, $b_{3}$, $m$, and $n$, so we consider specific values for the parameters.
The parameter constraint given Eq. (\ref{f2a}), with the following specific values $(n=2.5,m=-1)$ are depicted in Fig. \ref{fig:WECB2}; the constraints provided by Eq. (\ref{f2b}) are depicted in Fig. \ref{fig:WECB3} for the values $(n=1.8,m=-1)$; and finally the constraints presented by Eq. (\ref{f2c}) are depicted in Fig. \ref{fig:WECB4} for the values $(n=1.3,m=-1)$. The respective WEC conditions are then provided as a function of the parameters $a_3$ and $b_3$. As in the previous case, the plots depicted in Figs. \ref{fig:WECB2}-\ref{fig:WECB4} do indeed prove that the specific form of $ f_{2}(G)$ given by Eq. (\ref{terzo}) considered in \cite{Odintsov-models} is consistent with the WEC.

%%%%%%%%%%%%%%%%%%%%%%
\section{Discussion and final remarks}\label{ref:conclusion}
%%%%%%%%%%%%%%%%%%%%%%

The standard model of cosmology is remarkably successful in
accounting for the observed features of the Universe. However,
there remain a number of fundamental open questions at the
foundation of the standard model. In particular, we lack a
fundamental understanding of the acceleration of the late
universe. Recent observations of supernovae, together with the
WMAP and SDSS data, lead to the remarkable conclusion that our
universe is not just expanding, but has begun to
accelerate. One is liable to ask: What is the so-called `dark energy'
that is driving the acceleration of the universe? Is it a vacuum
energy or a dynamical field (``quintessence'')? Or is the
acceleration due to infra-red modifications of Einstein's theory
of General Relativity? How is structure formation affected in
these alternative scenarios? What will the outcome be of this
acceleration for the future fate of the universe?

The aspects of these fundamental questions whose resolution is so
important for theoretical cosmology, need to look beyond the
standard theory of gravity. A very promising way to explain these major problems is to assume that at large scales Einstein's theory of General Relativity breaks down, and a more general action describes the gravitational field. It is clear that these open questions involve not only gravity, but also particle physics. String theory provides a synthesis of these two parts of physics and is widely believed to be moving towards a viable quantum gravity theory. Thus, in considering alternative higher-order gravity theories, one is liable to be motivated in pursuing models consistent and inspired by several candidates of a fundamental theory of quantum gravity.
In this context, predictions of string/M-theory in the context of gravity-matter couplings, show that couplings of the scalar field with higher order curvature invariants are important. In particular, a coupling of the scalar field with the  Gauss-Bonnet invariant $G$ are fundamental in the appearance of non-singular early time cosmologies.

In this work, we discussed the viability of an interesting alternative gravitational theory, namely, modified Gauss-Bonnet gravity or $f(G)$ gravity. We considered specific realistic forms of $f(G)$ analyzed in the literature that account for the late-time cosmic acceleration and that cured the finite-time future singularities \cite{Nojiri:2007bt,Odintsov-models}. The general inequalities imposed by the energy conditions were outlined and using the recent estimated values of the Hubble, deceleration, jerk and snap parameters we have shown the viability of the above-mentioned forms of $f(G)$ imposed by the weak energy condition.
More specifically, for simplicity we only examined the vacuum case for which $p = \rho = 0$, although this is not a physically interesting case, as the universe contains matter 
and radiation.  However, this is easily corrected, since one can always 
add a positive energy density or pressure from matter and/or radiation 
to any model satisfying the WEC, and it will still satisfy the WEC. Thus, $f(G)$ gravity with matter will also satisfy the WEC if the vacuum model does.

However, as argued in \cite{Santos:2007bs} it is important to emphasize that although the energy conditions in modified theories of gravity have a well-founded physical motivation, i.e., the attractive nature of gravity as outlined in Raychaudhuri's equation, the issue as to whether they should be applied to modified theories of gravity is an open question, which is ultimately related to the confrontation between theory and observations.

\section*{Acknowledgments}
NMG acknowledges a postdoctoral fellowship from CONACYT-Mexico. The work of TH is supported by an RGC grant of the government of the Hong Kong SAR. FSNL and JPM acknowledge financial support of the Funda\c{c}\~{a}o
para a Ci\^{e}ncia e Tecnologia through the grants
PTDC/FIS/102742/2008 and CERN/FP/109381/2009.

%%%%%%%%%%%%%%%%%%%%%%%%%%%%

\end{document}